\documentclass[english]{iopart}
\usepackage[T1]{fontenc}
\usepackage[latin9]{inputenc}
\usepackage{booktabs}
\usepackage{amsbsy}
\usepackage{amstext}
\usepackage{amssymb}
\usepackage{graphicx}
\usepackage{esint}

\makeatletter

\providecommand{\tabularnewline}{\\}

\usepackage{iopams}
\usepackage{setstack}

\usepackage{braket}

\makeatother

\usepackage{babel}
\begin{document}
\title{Anomalous delocalization of resonant states in graphene \& the vacancy
magnetic moment }
\author{Mirko Leccese$^{1}$ \& Rocco Martinazzo$^{1,2,*}$}
\address{$^{1}$Department of Chemistry, Università degli Studi di Milano,
Via Golgi 19, 20133 Milano, Italy}
\address{$^{2}$Istituto di Scienze e Tecnologie Chimiche \textquotedblleft Giulio
Natta\textquotedblright , CNR, via Golgi 19, 20133 Milan, Italy}
\ead{rocco.martinazzo@unimi.it}
\begin{abstract}
Carbon atom vacancies in graphene give rise to a local magnetic moment
of $\sigma+\pi$ origin, whose magnitude is yet uncertain and debated.
Partial quenching of $\pi$ magnetism has been ubiquitously reported
in periodic \emph{first principles} calculations, with magnetic moments
scattered in the range 1.0 - 2.0 $\mu_{B}$, slowly converging to
the lower or the upper end, depending on how the diluted limit is
approached. By contrast, (ensemble) density functional theory calculations
on cluster models neatly converge to the value of $2$ $\mu_{B}$
when increasing the system size. This stunning discrepancy has sparked
an ongoing debate about the role of defect-defect interactions and
self-doping, and about the importance of the self-interaction-error
in the density-functional-theory description of the vacancy-induced
states. \\
Here, we settle this puzzle by showing that the problem has a fundamental,
mono-electronic origin which is related to the special (periodic)
arrangement of defects that results when using the slab-supercell
approach. Specifically, we report the existence of resonant states
that are \emph{anomalously} delocalized over the lattice and that
make the $\pi$ midgap band \emph{unphysically} dispersive, hence
prone to self-doping and quenching of the $\pi$ magnetism. Hybrid
functionals fix the problem by widening the gap between the spin-resolved
$\pi$ midgap bands, without reducing their artificial widths. As
a consequence, while reconciling the magnetic moment with expectations,
they predict a spin-splitting which is one order of magnitude larger
than found in experiments. 
\end{abstract}
\maketitle

\section{Introduction}

Defects in graphene such as monovalent ad-species or missing carbon
atoms - commonly referred to as ``$p_{Z}$ vacancies\textquotedblright{}
-- play important roles in charge transport, in magnetism and in
the chemistry of graphene, since they induce semi-localized \textquotedblleft midgap\textquotedblright{}
states which are known to decay slowly (as $1/r$) from the defect
position \cite{Pereira2006,Pereira2008,Ugeda2010}. These resonant
states host itinerant electrons, act as strong scatterers determining
graphene conductivity at both zero and finite carrier densities \cite{Robinson2008,CastroNeto2009,Wehling2010,Peres2010,Ferreira2011},
and provide spin-half semi-local magnetic moments which are detectable
by magnetometry experiments \cite{Nair2012,Nair2013} and bias graphene
chemical reactivity towards specific lattice positions \cite{HornekaerB2006,CasoloJCP2009,Bonfanti2018c}. 

Carbon atom vacancies play a special role in this respect since, besides
the above \textquotedblleft midgap\textquotedblright{} state, they
present additional dangling bonds in the $\sigma$ skeleton originating
from the removal of a C atom. In the bare vacancy a structural (Jahn--Teller)
distortion occurs \cite{El-Barbary2003,Ma2004,Lehtinen2004,Yazyev2007}
and leaves a lonely $\sigma$ electron which is free to couple with
the above mentioned $\pi$ one. As a result, the ground state of the
bare vacancy in large cluster models is predicted to be a (planar)
``triplet'', although a (non-planar) ``singlet'' with one spin
flipped (only \ensuremath{\sim}0.2 eV higher in energy) can be accessed
if ripples in the graphene sheet or interaction with a substrate are
taken into account \cite{Casartelli2013}. Periodic calculations do
predict a planar, high-spin configuration but fail in reproducing
the ``expected'' value of the magnetic moment \cite{Ma2004,Lehtinen2004,Yazyev2007,Palacios2012,Wang2012a,Casartelli2013}. 

It is well known that periodic calculations can be plagued by spurious
interactions between periodic images that determine the displacement
and the broadening of the defect-related bands, hence causing self-doping
and partial quenching of the magnetization. However, several works
have reported a \emph{reduction} of the local magnetic moments upon
increasing the cell size thereby suggesting a complete quenching of
the $\pi$ magnetism in the diluted limit \cite{Palacios2012}. Others
suggest an increase of the magnetic moment towards $\sim$1.5 $\mu_{B}$
when decreasing the defect concentrations \cite{El-Barbary2003,Ma2004,Lehtinen2004,Yazyev2007,Wang2012a,Casartelli2013},
and one of them predicts the ``expected'' limit of 2 $\mu_{B}$
when the limit of an isolated vacancy is carefully approached \cite{Rodrigo2016}.
As a matter of fact, the computed values of the magnetic moments scatter
in the range 1.0 - 1.7 $\mu_{B}$, and slowly converge to 1 $\mu_{B}$
\emph{or} 2 $\mu_{B}$, depending of how the limit of an isolated
vacancy is realized. By contrast, calculations based on clusters models
show that the magnetic moment increases with cluster size, clearly
tending to the expected value of 2 $\mu_{B}$ \cite{Wang2012a}. Therefore,
calculations including periodicity produce the opposite result of
cluster based simulations. 

This puzzling issue has been recently addressed by employing computationally
expensive hybrid functionals in periodic calculations, with the gratifying
reward of a reasonable value of the magnetic moment, smoothly converging
to $2$ $\mu_{B}$ when increasing the supercell size \cite{Valencia2017a,Ronchi2017}.
The immediate conclusion of these studies is that exact exchange is
crucial for the correct description of the vacancy magnetic moment:
hybrid functionals partially correct the self-interaction error (SIE)
that notoriously plagues local (LDA) and semi-local (GGA) functionals
and that can seriously affect the degree of electron localization/delocalization,
hence the magnetization. This seems to be a definite answer to the
long-debated issue of the magnetic moment of a C-atom vacancy in graphene,
but it is quite unsatisfactory from a theoretical point of view. Indeed,
this ``magnetic-moment problem'' does not involve at all the $\sigma$
electron\footnote{This is hosted in a well localized state featuring a strong Coulomb
repulsion ($\sim$ 2.5 eV) that prevents double occupation despite
the fact it is located well below the Fermi level ($\sim$ 0.7 eV). }, rather resides in the semilocalized (\emph{not} \emph{even normalizable})
``midgap'' state where \emph{e-e} interactions would hardly play
a significant role. All the more that GGA calculations for other $p_{Z}$
defects, e.g. H atoms, predict the ``correct'' $\pi$ moment, with
precisely the same defect-induced $\pi$ structure underneath \cite{Bonfanti2018c}. 

Here, motivated by these puzzling issues we reinforce some observations
made long ago by one of the present authors (R.M.), about the existence
of anomalously delocalized states in defective graphene that could
hamper the correct calculation of the magnetic moment in a periodic
setting \cite{Casartelli2013,Casartelli2014}. We do this by scrutinizing
the localization properties of the midgap state wavefunctions at both
the tight-binding and the \emph{first} \emph{principles} theory levels
(Section 3), after providing a comprehensive theoretical frame for
their discussion (Section 2) which should clarify the anomalous character
of the delocalization. By relating the anomalous delocalization of
the states to the \emph{dispersion} of the $\pi$ midgap band we provide
evidence for a spurious self-doping of the band, hence for an \emph{unphysical}
quenching of the magnetization. As we shall show below, the problem
is not limited to the C atom vacancy, although it affects the vacancy
more than other $p_{Z}$ defects. 

\section{Theory}

In this Section we present some background material which is necessary
to identify as \emph{anomalous }the behaviour of the midgap state
in a periodic calculation (at least in some regions of the supercell
Brilloin zone, SBZ). We start by providing some analytical results
about the expected spatial properties of the resonance induced by
a $p_{Z}$ vacancy in graphene, and then we show why -- at the tight
binding level with nearest-neighbors hoppings only --- \emph{fully}
delocalized states appear that contrast with the $1/r$ decaying states
expected for a vacancy. The delocalized solutions are inhereted from
the pristine system (they are ``robust'' against some disorder)
and, as will be shown numerically in Section 3, they survive at higher
level of theory and provide a mechanism for self-doping and for quenching
of the $\pi$ magnetization.

\subsection{Isolated $p_{Z}$ defect}

The most complete, yet analytical, description of the zero-mode state
induced by a $p_{Z}$ vacancy in graphene is provided by the non-interacting
Anderson model \cite{Anderson1961} for a simple adsorbed species
that covalently binds to one of the graphene lattice sites \cite{Robinson2008,Wehling2010}\footnote{See also Ref. \cite{Bonfanti2018c} for a detailed description of
the model and of the relevant results.}. The limit of a true C-atom vacancy can then be realized by strengthening
the adatom-lattice chemical bond up to the extent that the bonding
/ antibonding pair of molecular orbitals describing the bond separates
out from the lattice states (the so-called unitary limit). The ad-species
is described by a single level at energy $\epsilon_{\textrm{ad}}$
(e.g., the 1\emph{s} level of a H atom) which hybridizes with a carbon
atom of the lattice. If we place the latter at the origin of the lattice,
the total Hamiltonian reads as
\begin{equation}
H=H_{\text{latt}}+H_{\text{ad}}\label{eq:defective lattice H}
\end{equation}
where 
\begin{equation}
H_{\textrm{latt}}=-t\sum_{\sigma}\sum_{\braket{i,j}}c_{i,\sigma}^{\dagger}c_{j,\sigma}\label{eq:lattice H}
\end{equation}
is the tight-binding lattice Hamiltonian with nearest-neighbors hoppings
and
\begin{equation}
H_{\textrm{ad}}=\sum_{\sigma}\epsilon_{\textrm{ad}}d_{\sigma}^{\dagger}d_{\sigma}+W\sum_{\sigma}\left(c_{0,\sigma}^{\dagger}d_{\sigma}+d_{\sigma}^{\dagger}c_{0,\sigma}\right)\label{eq:H_adsorbate}
\end{equation}
Here, $W$ is the hybridization energy, $d_{\sigma}^{\dagger}$ $(d_{\sigma})$
creates (destroys) an electron with spin $\sigma$ in the adatom energy
level, $c_{i,\sigma}^{\dagger}$ ($c_{i,\sigma}$) does the same for
the lattice site $i$ and $t$ is the hopping energy for the nearest-neighbors
pairs $\braket{i,j}$. We are interested in the Green's operator\footnote{As usual, $\lambda$ is understood to be $\lambda=\epsilon+i\eta$,
where $\epsilon$ is a real energy and $\eta\rightarrow0^{+}$.} $G(\lambda)$ of the one-electron Hamiltonian, which can be obtained
upon partitioning the one-electron space into a primary subspace (the
lattice) and the remainder \cite{Levine1969}. Its projection onto
the lattice --- i.e. $PG(\lambda)P$, $P$ being the projector onto
the primary subspace --- takes the form $PG(\lambda)P=(\lambda-H_{\textrm{eff}})^{-1}\equiv G_{\text{eff}}(\lambda)$
where $H_{\textrm{eff}}$ is an effective, energy-dependent Hamiltonian
implicitly accounting for the dynamics in the adatom level. The latter
reads as $H_{\textrm{eff}}=H_{\textrm{latt}}+V(\lambda)$ with the
local scattering potential defined by
\[
V(\lambda)=W^{2}\sum_{\sigma}\frac{\ket{\chi_{0,\sigma}}\bra{\chi_{0,\sigma}}}{\lambda-\epsilon_{\textrm{ad}}}
\]
where $\ket{\chi_{0,\sigma}}$ is a $p_{Z}$ spin-orbital at the defective
site. The sought for effective Green function can be obtained from
the $T$ matrix \cite{Taylor1969}, $T(\lambda):=V+VG_{\textrm{eff}}(\lambda)V$,
since the corresponding Lippmann-Schwinger equation $T(\lambda)=V+VG^{0}(\lambda)T(\lambda)$
(where $G^{0}$ is the Green's operator of the unperturbed lattice)
is solved by 
\begin{equation}
T(\lambda)=t(\lambda)\sum_{\sigma}\ket{\chi_{0,\sigma}}\bra{\chi_{0,\sigma}}\ \ \ \ t(\lambda)=\frac{W^{2}}{\lambda-\epsilon_{\textrm{ad}}-W^{2}g_{00}^{0}(\lambda)}\label{eq:T-matrix}
\end{equation}
when the potential takes a separable form. Here, $g_{00}^{0}(\lambda)$
is the on-site Green's function of the unperturbed lattice, 
\[
g_{00}^{0}(\epsilon)=\braket{\mathbf{0}\sigma|G^{0}(\lambda)|\mathbf{0}\sigma}\approx-\frac{\epsilon}{\epsilon_{c}^{2}}\ln\left(\left|\frac{\epsilon_{c}^{2}}{\epsilon^{2}}-1\right|\right)-i\pi\rho^{0}(\epsilon),
\]
$\rho^{0}(\epsilon)\approx\frac{|\epsilon|}{\epsilon_{c}^{2}}\Theta(\epsilon_{c}-|\epsilon|)$
is the density of states per C atom per spin channel and $\epsilon_{c}$
is an energy cutoff that determines the bandwidth, $\epsilon_{c}\approx\hbar v_{F}d_{\text{CC}}^{-1}$,
where $d_{\text{CC}}\approx1.42\ $\AA$\ $ is the C-C bond length\footnote{Upon matching $\rho^{0}(\epsilon$) to the known low-energy expression
for the density of states in graphene one obtains $k_{c}=\epsilon_{c}/\hbar v_{F}=2\sqrt{2\pi/3\sqrt{3}}d_{CC}^{-1}\approx2d_{CC}^{-1}$
or $\epsilon_{c}=t\sqrt{\pi\sqrt{3}}\approx6$ eV.}. Because of the linear density of states, the real part of the on-site
Green's function features a vertical cusp at zero energy that has
a huge impact on the defect-induced states. Indeed, it makes the position
of the resonance defined by Eq. \ref{eq:T-matrix}, i.e. the lowest
energy solution of the equation 
\[
\epsilon-\epsilon_{\textrm{ad}}-W^{2}\Re g_{00}^{0}(\epsilon)=0,
\]
rather insensitive to the hybridization energy $W$ and always very
close to $\epsilon=0$. This feature gives an ``universal'' character
to the defect problem, meaning that it does not depend on the details
of the defect originating the $p_{Z}$ vacancy in the lattice. 

We are interested in the spatial properties of such resonance, i.e.
in the scattered component $\ket{\psi_{\epsilon}^{\textrm{scatt}}}$
of the scattering state $\ket{\psi_{\epsilon}+}$, 
\[
\ket{\psi_{\epsilon}+}=\ket{\psi_{\epsilon}^{0}}+\ket{\psi_{\epsilon}^{\textrm{scatt}}}=\ket{\psi_{\epsilon}^{0}}+G^{0}(\epsilon)T(\epsilon)\ket{\psi_{\epsilon}^{0}}
\]
where $\ket{\psi_{\epsilon}^{0}}$ is an eigenstate of the unperturbed
lattice with energy $\epsilon$. Clearly, 
\[
\psi_{\epsilon}^{\textrm{scatt}}(\mathbf{r})\propto t(\epsilon)G^{0}(\mathbf{r},\mathbf{0}|\epsilon)
\]
where $G^{0}(\mathbf{r},\mathbf{0}|\epsilon)=\braket{\mathbf{r}\sigma|G^{0}(\epsilon)|\mathbf{0}\sigma}$
is the Green function of the unperturbed lattice at $\mathbf{r}$
for the adatom sitting on the site at the origin $\mathbf{0}$ (which
we take to be of A type in the following). This quantity has been
considered by several authors for different reasons --- from STM
imaging \cite{Wang2006} to RKKY interactions in graphene \cite{SherafatiA2011,SherafatiB2011,Kogan2011}
--- and can be obtained numerically as Fourier transform of the (simpler)
Green's function in $k-$space, namely from 
\[
G_{\text{XA}}^{0}(\mathbf{r},\mathbf{0}|\epsilon)=\frac{1}{\Omega_{\text{BZ}}}\int_{\text{BZ}}d^{2}\mathbf{k}e^{i\mathbf{k}\mathbf{r}}G_{XA}^{0}(\mathbf{k}|\epsilon)
\]
where X=A,B depending on whether $\mathbf{r}$ is a lattice position
in the A or B sublattice ($\mathcal{A}$ and $\mathcal{B}$, respectively,
in the following). Here, the integral runs over the graphene Brillouin
zone (BZ) and $\Omega_{\text{BZ}}$ is its area. At the low energies
of interest for the defect-induced resonance, the above expression
can be integrated analytically in the linear band approximation with
negligible error for not too small distances ($r>d_{\text{CC}}$).
The result is
\begin{equation}
G_{\text{AA}}^{0}(\mathbf{r},\mathbf{0}|\epsilon)=-i\frac{A_{c}}{2}\frac{|\epsilon|\cos(\mathbf{K}\mathbf{r})}{\hbar^{2}v_{F}^{2}}H_{0}^{\pm}\left(\frac{|\epsilon|r}{\hbar v_{F}}\right)\ \ \ \ \textrm{for}\ \mathbf{r}\in\mathcal{A}\label{eq:G_AA}
\end{equation}
\begin{equation}
G_{\text{BA}}^{0}(\mathbf{r},\mathbf{0}|\epsilon)=+i\frac{A_{c}}{2}\frac{\epsilon\sin(\mathbf{K}\mathbf{r}+\theta)}{\hbar^{2}v_{F}^{2}}H_{1}^{\pm}\left(\frac{|\epsilon|r}{\hbar v_{F}}\right)\ \ \ \ \textrm{for}\ \mathbf{r}\in\mathcal{B}\label{eq:G_BA}
\end{equation}
where $H_{l}^{\pm}$ are Hankel functions of the first and second
kind (for positive and negative energies, respectively) of order $l$,
$A_{c}$ is the area of the graphene unit cell and $\theta$ is the
angle between $\mathbf{r}$ and vector $\mathbf{K}$. The latter locates
a K corner in the BZ and can be given as $2\pi\mathbf{K}=\Omega_{\text{BZ}}\boldsymbol{\delta}\wedge\hat{\mathbf{n}}$,
where $\boldsymbol{\delta}$ is the position of the B site relative
to the A one in the (arbitrarily) chosen unit cell, and $\hat{\mathbf{n}}$
is the surface normal. Thus, the scattering resonance turns out to
have the expected threefold symmetry, with maxima along the armchair
directions. It further presents the celebrated $1/r$ decay in an
intermediate distance range $d_{\text{CC}}<r\ll\hbar v_{F}/|\epsilon|$,
that can be effectively rather wide at the energies of interest ($\epsilon\rightarrow0$).
Indeed, upon replacing the Hankel's functions in Eqs. \ref{eq:G_AA}
and \ref{eq:G_BA} with their low-argument expansion one obtains the
approximate expressions
\begin{equation}
G_{\text{AA}}^{0}(\mathbf{r},\mathbf{0}|\epsilon)\approx\frac{A_{c}}{\pi}\frac{\epsilon r}{\hbar v_{F}}\frac{\cos(\mathbf{K}\mathbf{r})}{\hbar v_{F}r}\left[\ln\left(\frac{|\epsilon|r}{2\hbar v_{F}}\right)+\gamma\right]\ \ \ \ \textrm{for}\ \mathbf{r}\in\mathcal{A}\label{eq:G_AA - intermediate range}
\end{equation}
\begin{equation}
G_{\text{BA}}^{0}(\mathbf{r},\mathbf{0}|\epsilon)\approx\frac{A_{c}}{\pi}\frac{\sin(\mathbf{K}\mathbf{r}+\theta)}{\hbar v_{F}r}\ \ \ \ \textrm{for}\ \mathbf{r}\in\mathcal{B}\label{eq:G_BA - intermediate range}
\end{equation}
(where $\gamma\approx0.577$ is the Eulero-Mascheroni constant) featuring
a $1/r$ decay of $G^{0}$ on the majority sublattice ($\mathcal{B}$)
and a much smaller magnitude on the minority one ($\mathcal{A}$),
vanishing for $\epsilon\rightarrow0$\footnote{For the minority sublattice contribution notice that, with $x=|\epsilon|r/\hbar v_{F}$,
one has $|x\ln x|\leq e^{-1}$ for any $x<1$.}. Such algebraic decay of the wavefuncion was first predicted \cite{Cheianov2006,Pereira2006,Mariani2007,Pereira2008,Bena2008}
and experimentally observed \cite{Ugeda2010} for carbon atom vacancies
and more recently investigated for hydrogen atoms \cite{Gonzalez-Herrero2016}.
The above equations provide a full spatial description of the $\pi$
midgap state, which is valid at any distance from the defect position
but the smallest ones (i.e., for $r>d_{\text{CC}}$), which however
are irrelevant for the present discussion. 

\subsection{Periodic arrangement of $p_{Z}$ defects }

Next we consider the case --- which is rather common when it comes
to modeling with \emph{first} \emph{principles} means -- where the
$p_{Z}$ defect is investigated in a periodic setting with a sufficiently
large supercell that minimizes any electronic / structural interaction
between the defect and its periodic image. We consider again a tight
binding model with next-neighbors hoppings only and show that, irrespective
of the supercell shape and size there exist regions in the superlattice
Brillouin zone where the zero energy modes are anomalous, i.e. they
fully delocalize over the lattice \emph{without} decaying as $1/r$
away from the defect positions, as it happens for the isolated defect.
These anomalous states do not disperse at the simple tight binding
level of theory but do it in the presence of inhomogeneities in the
lattice, thereby affecting the correct occupation of $k-$states in
the Brillouin zone and the net induced magnetic moment. 

We start observing that, at the above mentioned tight binding level,
one of the two linearly independent states that are found at the $K$
point of pristine graphene is also an exact zero-energy solution for
the lattice with a $p_{Z}$ vacancy (the same happens at the $K'$
point). To this end we re-write the lattice Hamiltonian of the unperturbed
system, Eq. \ref{eq:lattice H}, by explicitly showing its bipartite
nature
\[
H_{\text{latt}}=-t\sum_{\sigma}\sum_{\boldsymbol{R},i}a_{\boldsymbol{R},\sigma}^{\dagger}b_{\boldsymbol{R}+\boldsymbol{\delta}_{i},\sigma}+\text{c.c.}
\]
Here the inner sum runs over the lattice sites, $\boldsymbol{\delta}_{i}$
are the three vectors joining $A$ sites to $B$ sites, and $a_{\boldsymbol{R},\sigma}$,
$b_{\boldsymbol{R},\sigma}$ are the previous $c_{i,\sigma}$'s annihilation
operators, now for the $A$ and $B$ sites at $\boldsymbol{R},$ respectively.
The Hamiltonian for the defective lattice with a missing $A$ site
at the origin $\boldsymbol{R}=\mathbf{0}$ then reads as 
\begin{equation}
H=H_{0}+t\sum_{\sigma}\sum_{i=1}^{3}\left(a_{\boldsymbol{0},\sigma}^{\dagger}b_{\boldsymbol{\delta}_{i},\sigma}+b_{\boldsymbol{\delta}_{i},\sigma}^{\dagger}a_{\boldsymbol{0},\sigma}\right)\label{eq:tight-binding Hamiltonian with vacancy}
\end{equation}
In pristine graphene, the eigenstates at the $K$ (or $K'$) point,
because of degeneracy, can be chosen to localize on either sublattice,
i.e. to have definite projection of the pseudo-spin along $z$. Let
$\ket{\phi_{K,\sigma}^{\text{A}}}$ and $\ket{\phi_{K,\sigma}^{\text{B}}}$
be such localized states at the $K$ point, respectively for $\mathcal{A}$
and $\mathcal{B}$. Since they are eigenstates at zero energy we have
\[
\sum_{i=1}^{3}a_{\boldsymbol{R},\sigma}^{\dagger}b_{\boldsymbol{R}+\boldsymbol{\delta}_{i},\sigma}\ket{\phi_{K,\sigma}^{\text{B}}}=0
\]
for any lattice vector $\boldsymbol{R}$, hence it holds
\[
\sum_{i=1}^{3}\left(a_{\boldsymbol{0},\sigma}^{\dagger}b_{\boldsymbol{\delta}_{i},\sigma}+b_{\boldsymbol{\delta}_{i},\sigma}^{\dagger}a_{\boldsymbol{0},\sigma}\right)\ket{\phi_{K,\sigma}^{\text{B}}}\equiv0
\]
This shows that $\ket{\phi_{K,\sigma}^{\text{B}}}$ is also an eigenstate
of the defective Hamiltonian $H$ at the same energy. 

Next, we observe that that this property is not limited to a single
vacancy, since similar reasoning applies to an \emph{arbitrary} number
of isolated vacancies, provided they are all placed in the $A$ sublattice.
In particular, it holds for a \emph{periodic} arrangement of vacancies
in large supercells, which is the case of interest when modeling $p_{Z}$
vacancies with a periodic-supercell approach. In such instance, wherever
the $K$ and $K'$ points fold to into the SBZ there appear `preserved'
states of the pristine lattice that occupy the majority lattice sites
only (as the midgap state ought to) but have an extended character,
without decaying as $1/r$ away from the defect. And, by continuity,
it is expected on general grounds that this anomalous delocalization
of the wavefunction is not limited to the above special points, rather
it extends over their neighborhoods in the SBZ. This feature ---
while not affecting the dispersion of the midgap $\pi$ \emph{band}
for the (periodized) Hamiltonian of Eq. \ref{eq:tight-binding Hamiltonian with vacancy}\footnote{This remains flat all over the SBZ since it is constrained by $e-h$
symmetry to lie at zero energy. }--- do affect the properties of the corresponding band when, more
realistically, $e-h$ symmetry breaks down. For instance, a slight
change of the on-site energy has to be expected around a C atom vacancy
because of the charge re-distribution that unavoidably occurs at an
edge of a graphenic structure, and this suggests that the midgap $\pi$
band becomes dispersive in those regions of the SBZ where the states
are delocalized. In turn, this affects the occupation of the band
and the net magnetic moment that it carries. We stress that this dispersion
is \emph{unphysical, }meaning that it is a feature of the periodic
arrangement of vacancies rather than the system one would (most probably)
like to describe, i.e. an isolated vacancy in graphene. 

\section{Results}

In this Section we present some numerical results that support the
above interpretation that the dispersion of the $\pi$ midgap band
in a periodic model of vacancies is unphysical (for the purpose of
describing an isolated vacancy in the lattice). We start by considering
simple tight-binding (TB) calculations in which we modelled the vacancy
at a very simple level but using relatively large supercells. Then
we move to a more realistic description of the C-atom vacancy in graphene
by considering density functional theory (DFT) calculations that account
for the atomistic details of the defect in the lattice, albeit in
a more limited setting. 

\subsection{Tight-binding}

We first investigate the midgap $\pi$ band in periodic TB Hamiltonians
with $p_{Z}$ vacancies, using unperturbed (i.e., homogeneous) on-site
energies. We consider for simplicity $n\times n$ superlattices where
the anomalies are easily identified since, depending on the superlattice
constant $n$, $K$ and $K'$ fold either to $K/K'$ or to the $\Gamma$
point of the SBZ. Specifically, when $n\ne3m$ ($m\in\mathbb{N}$)
\emph{non-degenerate} graphene zero-energy modes are expected at both
the $K$ and the $K'$ point of the SBZ, since $K$ and $K'$ fold,
respectively, into $K$ and $K'$ for $n=3m+1$ and into $K'$ and
$K$ for $n=3m+2$. When $n=3m$, on the other hand, both $K$ and
$K'$ fold into $\Gamma$, a Dirac cone survives, and three states
are expected at zero energy, two of which (the ones localizing on
the majority sites) are necessarily the extended $K$ and $K'$ states
of pristine graphene. 
\begin{figure}
\begin{centering}
\includegraphics[clip,width=0.9\columnwidth]{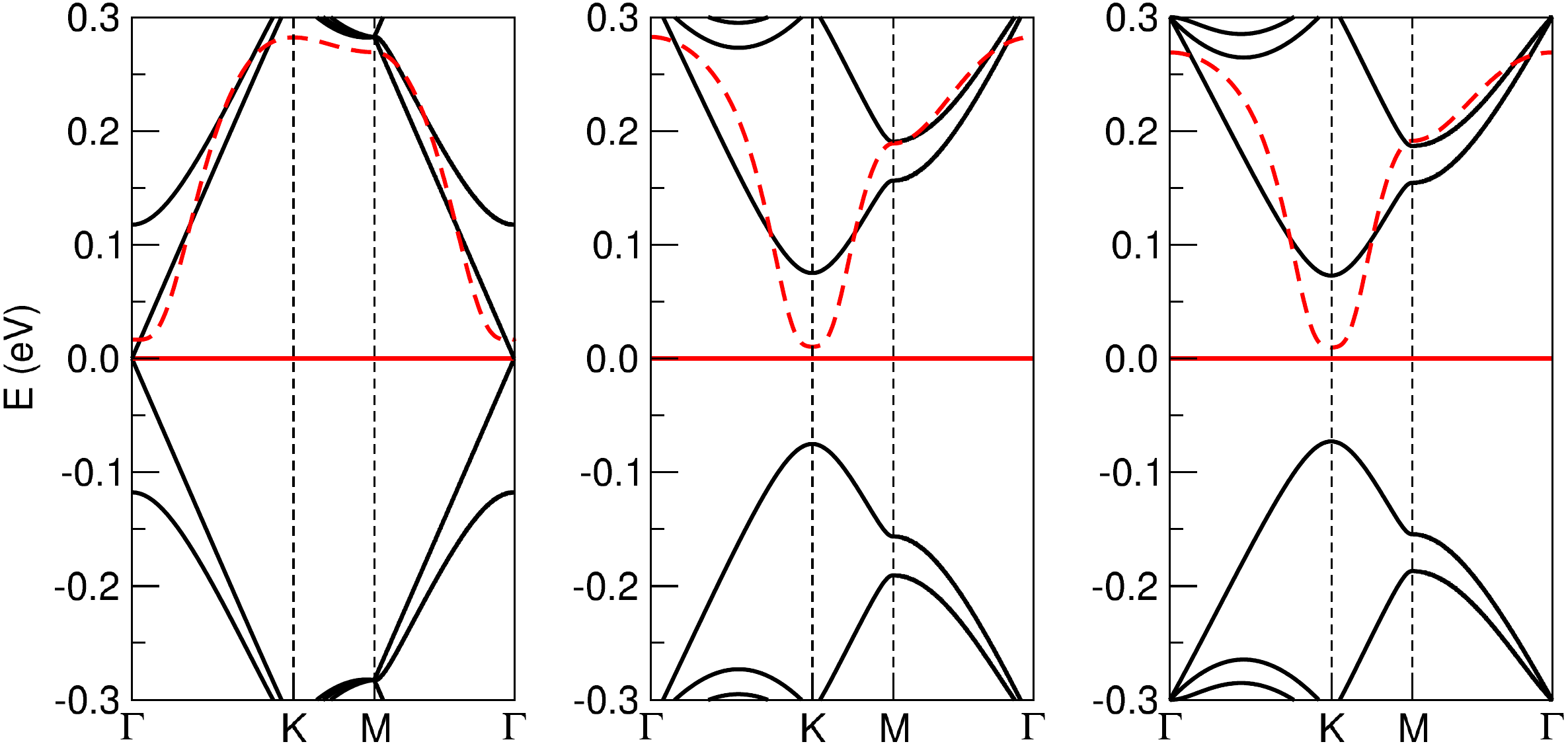}
\par\end{centering}
\caption{\label{fig:Low-energy-tight-binding}Low energy band structure of
$n\times n$ superlattices with one vacancy from tight-binding calculations
with $t=2.7$ eV. $n=30,31,32$ from left to right. Full red lines
for the midgap band and dashed red lines for the corresponding inverse
participation ratio, see text for details. For $n=30$ the latter
was computed for a state localized on the majority sublattice.}
\end{figure}

\begin{figure}
\begin{centering}
\includegraphics[clip,width=0.9\textwidth]{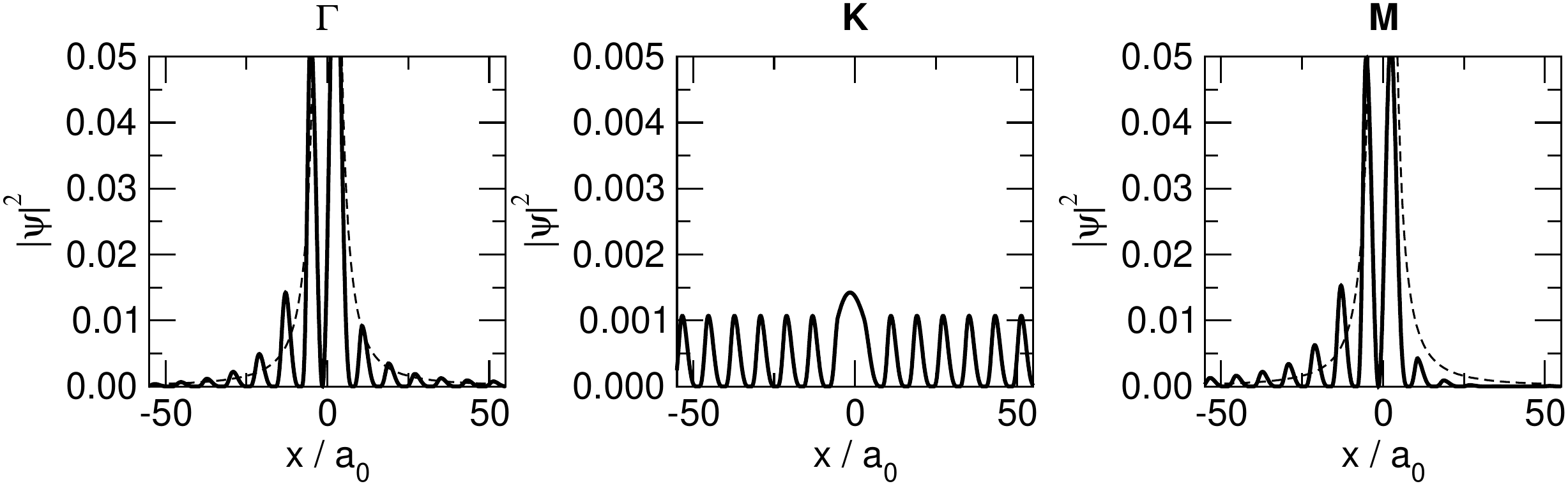}
\par\end{centering}
\caption{\label{ZEM2}Probability density of the zero-energy modes found at
the special points of the superlattice Brillouin zone for $n=31$,
from left to right for $\Gamma,K$ and $M$, as indicated. The $x$
axis is placed along one of the three-fold symmetry direction of the
defective lattice. Curves are Akima splines interpolating the numerically
determined coefficients and dashed lines on the leftmost and on the
rightmost panels represent the functions $\pm1/r^{2}$.}
\end{figure}

Fig. \ref{fig:Low-energy-tight-binding} shows the band-structure
of three different superlattices with a (super)lattice constant corresponding
to a rather large unit cell, $n=30,31,32$ from left to right. In
these calculations the hopping parameter was given the value $t=2.7\ $
eV to ease the comparison with the \emph{first principles} results
discussed below. In Fig. \ref{fig:Low-energy-tight-binding} the non-dispersive
midgap band is plotted in red (full line) and it is accompanied (dashed
red line, right scale) by its $q=2$ inverse participation ratio,
which is defined as
\begin{equation}
\text{IPR}_{q}=\sum_{i}|\psi_{i}|^{2q}\label{eq:IPR (discrete)}
\end{equation}
where the sum runs over the lattice sites of the supercell and $\psi_{i}$
is the amplitude of the zero-energy mode at the $i^{\text{th}}$ site.
This IPR$_{q}$ (for $q>1$) measures the localization of the midgap
states in this periodic arrangement of defects. A (normalized) state
which is fully delocalized over the lattice gives $\text{IPR}_{2}=1/n^{2}$
, since $n^{2}$ is the number of majority sites in a $n\times n$
supercell where the midgap state is known to localize. For a state
decaying as $1/r$, on the other hand, the inverse participation ratio
takes a larger value, namely $\text{IPR}_{2}\approx a/(b+\mbox{ln}n)^{2}$
where $a$ and $b$ are constants and the dependence on $n$ follows
from the logarithmic divergence of its squared norm. Clearly, the
real-space localization properties of the midgap state change across
the SBZ and the state does indeed delocalize where the special $K$
and $K'$ points of graphene are found to fold in the SBZ. Fig. \ref{ZEM2}
shows the details of the zero energy wavefunctions for $n=31$ along
one of the three-fold symmetry directions of the vacant lattice, and
explicitly proves the existence of delocalized states at special symmetry
points of the superlattice Brilloiun zone. Similar results are found
for $n=30,32$, with the only \emph{caveat} that for $n=30$ three
delocalized states are found at $\Gamma$ as a consequence of the
residual degeneracy. 

\begin{figure}
\noindent \begin{centering}
\includegraphics[clip,width=0.7\columnwidth]{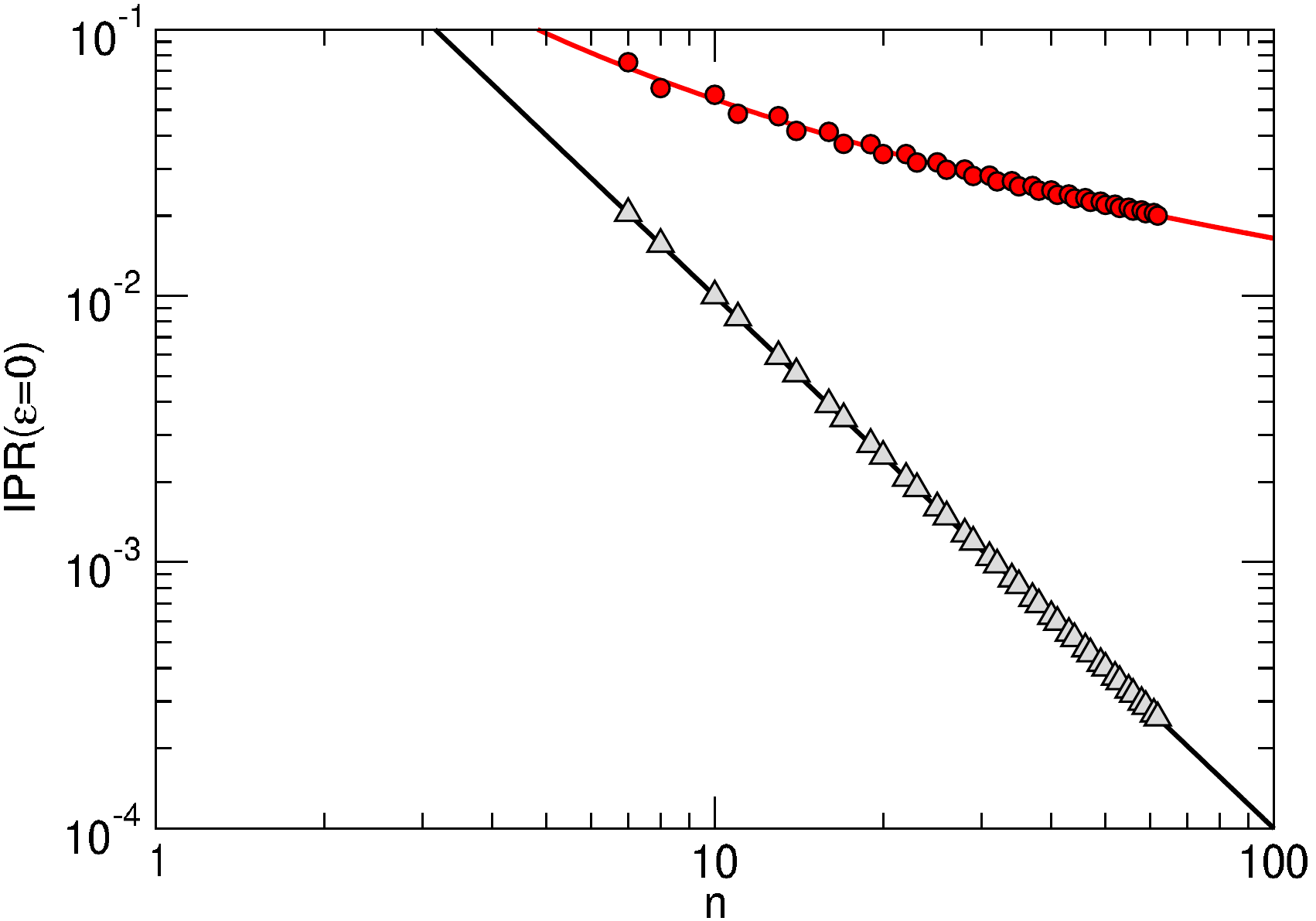}
\par\end{centering}
\caption{\label{fig:IPR}Inverse participation ratio for anomalous (grey triangles)
and ordinary (red circles) midgap states in $n\times n$ supercells,
as computed at the $K$ and $\Gamma$ points of the superlattice Brillouin
zones, respectively, for $n$ not a multiple of $3$. Also shown as
full lines, the corresponding theoretical values, namely $\text{IPR}_{2}=1/n^{2}$
for a fully delocalized state (black curve) and a fit to $\text{IPR}_{2}\approx a/(b+\mbox{ln}n)^{2}$
appropriate for a state decaying as $1/r$ (red curve). }
\end{figure}

These odd findings are not limited to the above (already large) unit
cells, rather persist for any supercell size. This is seen by the
$\text{IPR}_{2}$ computed at the $K$ and the $\Gamma$ points for
several $n\times n$ superlattices of $p_{Z}$ vacancies, which is
reported in Fig.\ref{fig:IPR}. {[}Please notice that in drawing the
figure we set $n\neq3m$ to free ourselves from the annoying degeneracy
problem that prevents the automatic analysis of the zero energy mode{]}.
Also shown in the same figure (full black lines) is the expected behavior
of the IPR$_{2}$, that confirms the rather different spatial extensions
of the states at the selected special points of the SBZ.

\begin{figure}
\noindent \begin{centering}
\includegraphics[clip,width=0.9\columnwidth]{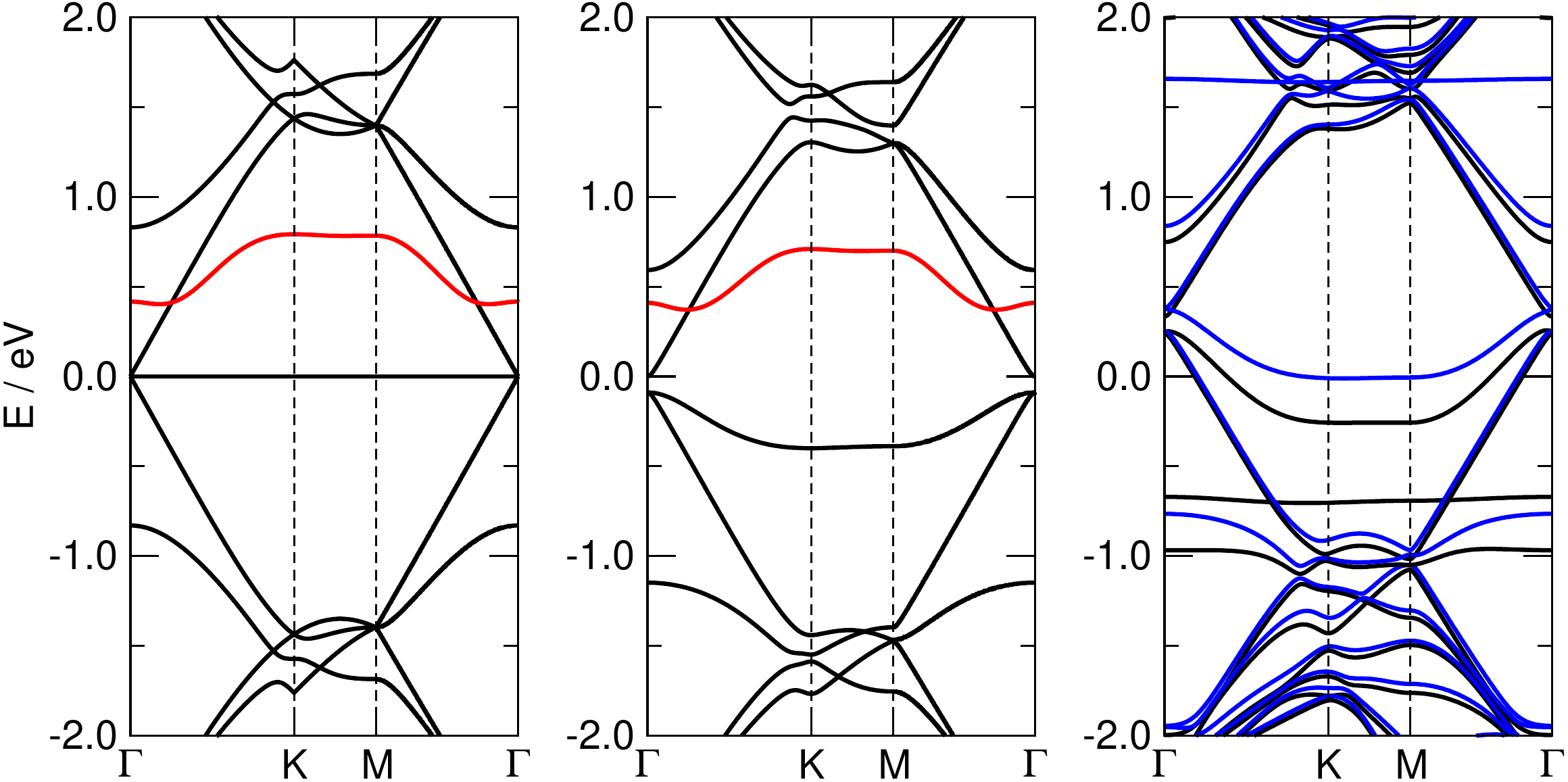}
\par\end{centering}
\caption{\label{fig:Band-structures-of}Band structure of a carbon atom vacancy
in a $6\times6$ supercell. Left and middle panels: energy bands (black
curves) from tight binding calculations with hopping parameter $t=2.7$
eV. Red curves are for the inverse participation ratio of the midgap
state. Calculations in the middle panel used an energy offset of $\Delta\epsilon=-1.0$
eV for the on-site energy of the three sites closest to the vacancy.
Right panel: spin-polarized \emph{first principles} calculations,
with black and blue curves for majority spin and minority spin states,
respectively. Energies are referenced to the Fermi level.}
\end{figure}

Finally, we illustrate how a varying real-space localization may turn
a flat band to be dispersive in the realistic situation where some
inhomogeneities in the lattice are present. Fig. \ref{fig:Band-structures-of}
shows the band structure of a $p_{Z}$ vacancy in a $6\times6$ cell,
as obtained at different levels of theory: tight binding for $\pi$
states only in the left and middle panels, and \emph{first principles}
calculations (to be described below) for a C-atom vacancy in the rightmost
panel. For the left panel, tight binding calculations used (as above)
a uniform distribution of on-site energies, while for the middle panel
we introduced some inhomogeneity in the lattice. Specifically, we
shifted the on-site energy of the three lattice sites closest to the
defect by $1.0$ eV below the value it takes in pristine graphene,
in such a way to mimic the charge redistribution occurring in the
$\sigma$ bands as a consequence of the removal of a C atom. The figure
makes clear that the anomalously delocalized zero energy state that
appears close to $\Gamma$ leads, in the presence of inhomogeneities,
to a dispersive migdap band which (with the chosen parameters) closely
resembles the one found from \emph{first} \emph{principles} in the
atomistic model of the vacancy. At the \emph{first} \emph{principles}
level of theory self-doping necessarily occurs, the Dirac point shifts
above the Fermi level and the spin-up and spin-down bands split. Hence,
the majority spin channel is only fractionally occupied and the magnetization
gets partially quenched, if the comparison is made with the situation
of a spin-split but otherwise flat band. We stress that the ``dispersive
region of the SBZ'' shrinks together with the superlattice Brillouin
zone when increasing the supercell size \emph{without} disappearing.
Hence, in self-consistent calculations, in order to obtain a smooth
behavior of the magnetic moment with the system size and to attempt
any kind of extrapolation, the larger the supercell is the finer the
$k$ mesh should be, as indeed observed in calculations \cite{Rodrigo2016}.

\subsection{First principles calculations}

Next, we consider the results of several \emph{first} \emph{principles}
calculations that we performed on different periodic arrangements
of C atom vacancies and similar $p_{Z}$ vacancies. In these calculations
we used the pseudo-potential density functional theory implementation
available in SIESTA \cite{SIESTA,SIESTA2}, in the generalized gradient
approximation provided by the Perdew-Burke-Ernzerhof functional. Separable
\cite{KBSeparable} norm conserving pseudopotentials \cite{TMPseudo}
with partial core corrections \cite{PPCoreCorrection} were used to
replace core electrons, and Kohn-Sham states were represented on a
basis of numerical atomic orbitals with compact support. Such an approach
allows one to tackle efficiently huge-sized problems, at the expense
of a reduced control on the convergence of the one-electron basis.
For this reason we used consistently a double-$\zeta$ basis set with
single polarization orbitals (DZP) but occasionally checked the results
with triple-$\zeta$ sets and double polarization orbitals (TZ2P),
finding negligible differences between the two sets of results. We
optimized each structure considered adopting a stringent threshold
on the maximum component of the atomic forces (0.005 eV/\AA ), in
conjunction with a large energy cutoff (800 Ry) for the real space
integration grid to remove any egg-box effect.
\begin{table}
\begin{centering}
\begin{tabular}{ccccc}
\toprule 
Supercell & $k$ - mesh & $d_{<}/$\AA & $d_{>}/$\AA & $M/\mu_{B}$\tabularnewline
\midrule
5x5 & 4x4 & 2.128 & 2.552 & 1.767\tabularnewline
 & 5x5 & 2.132 & 2.553 & 1.820\tabularnewline
 & 6x6 & 2.123 & 2.551 & 1.720\tabularnewline
 & 8x8 & 2.127 & 2.552 & 1.752\tabularnewline
 & 10x10 & 2.129 & 2.552 & 1.767\tabularnewline
 & 12x12 & 2.127 & 2.552 & 1.751\tabularnewline
 & 24x24 & 2.130 & 2.553 & 1.762\tabularnewline
\midrule 
6x6 & 4x4 & 2.073 & 2.551 & 1.673\tabularnewline
 & 5x5 & 2.070 & 2.549 & 1.667\tabularnewline
 & 6x6 & 2.067 & 2.548 & 1.671\tabularnewline
 & 8x8 & 2.066 & 2.548 & 1.638\tabularnewline
 & 10x10 & 2.069 & 2.548 & 1.654\tabularnewline
 & 12x12 & 2.069 & 2.548 & 1.654\tabularnewline
 & 24x24 & 2.069 & 2.549 & 1.654\tabularnewline
\midrule 
7x7 & 4x4 & 2.026 & 2.544 & 1.637\tabularnewline
 & 5x5 & 2.012 & 2.542 & 1.491\tabularnewline
 & 6x6 & 2.016 & 2.543 & 1.562\tabularnewline
 & 8x8 & 2.015 & 2.543 & 1.547\tabularnewline
 & 10x10 & 2.016 & 2.543 & 1.545\tabularnewline
 & 12x12 & 2.015 & 2.542 & 1.540\tabularnewline
 & 24x24 & 2.014 & 2.543 & 1.535\tabularnewline
\bottomrule
\end{tabular}
\par\end{centering}
\caption{\label{tab: convergence monovacancy}Short ($d_{<}$) and long ($d_{>}$)
carbon-carbon distances in the monovacancy and corresponding magnetic
moment $M$ from the \emph{first-principles} calculations described
in the main text, for different supercell size and $k$-mesh. For
comparison, plane wave calculations for the 6x6 supercell 6x6 with
a 6x6 $k$-mesh result in $d_{<}=2.007$ \AA , $d_{>}=2.557$ \AA
, $M=1.556\ \mu_{B}$.}
\end{table}

The detailed structure of a C-atom vacancy is well-known and well
described in the literature. Briefly, the removal of a C atom creates
a $p_{Z}$ defect responsible for a $\pi$ resonant state and, at
the same time, leaves three dangling orbitals in the $\sigma$ network.
In the $D_{3h}$ point symmetry group of the system the first belongs
to the $a_{2}''$ symmetry species, while the second span the $a_{1}'+e'$
irreducible representations. Among the latter, $a'_{1}$ is lowest
in energy since it contains a purely bonding combination of $\sigma$
orbitals. Hence, the lowest energy configuration of the many-body
state is of $E''$ symmetry and undergoes a standard $E\otimes e$
Jahn-Teller distortion driven by the coupling with in-plane $e'$
vibrations. As a result, a distorted geometry with a \textquotedblleft pentagonal\textquotedblright{}
ring and an \textquotedblleft apical\textquotedblright{} carbon atom
opposite to it emerges as equilibrium configuration, threefold degenerate.
The \emph{vertical} position of the apical C atom determines the preferred
spin alignment (through a second order vibronic coupling): in the
lowest energy state, this is the high-spin state with the apical C
atom in-plane with the graphene sheet. Periodic DFT calculations correctly
predict the in-plane arrangement but present a difficult convergence
for the monovacancy properties, particularly for the magnetization,
as exemplified by Table \ref{tab: convergence monovacancy} that reports
some details for the smallest supercells considered. As we now show,
this is due to the anomalous delocalization and the ensuing unphysical
dispersion of the midgap-state $\pi$ band.

\begin{figure}
\begin{centering}
\includegraphics[width=0.6\paperwidth]{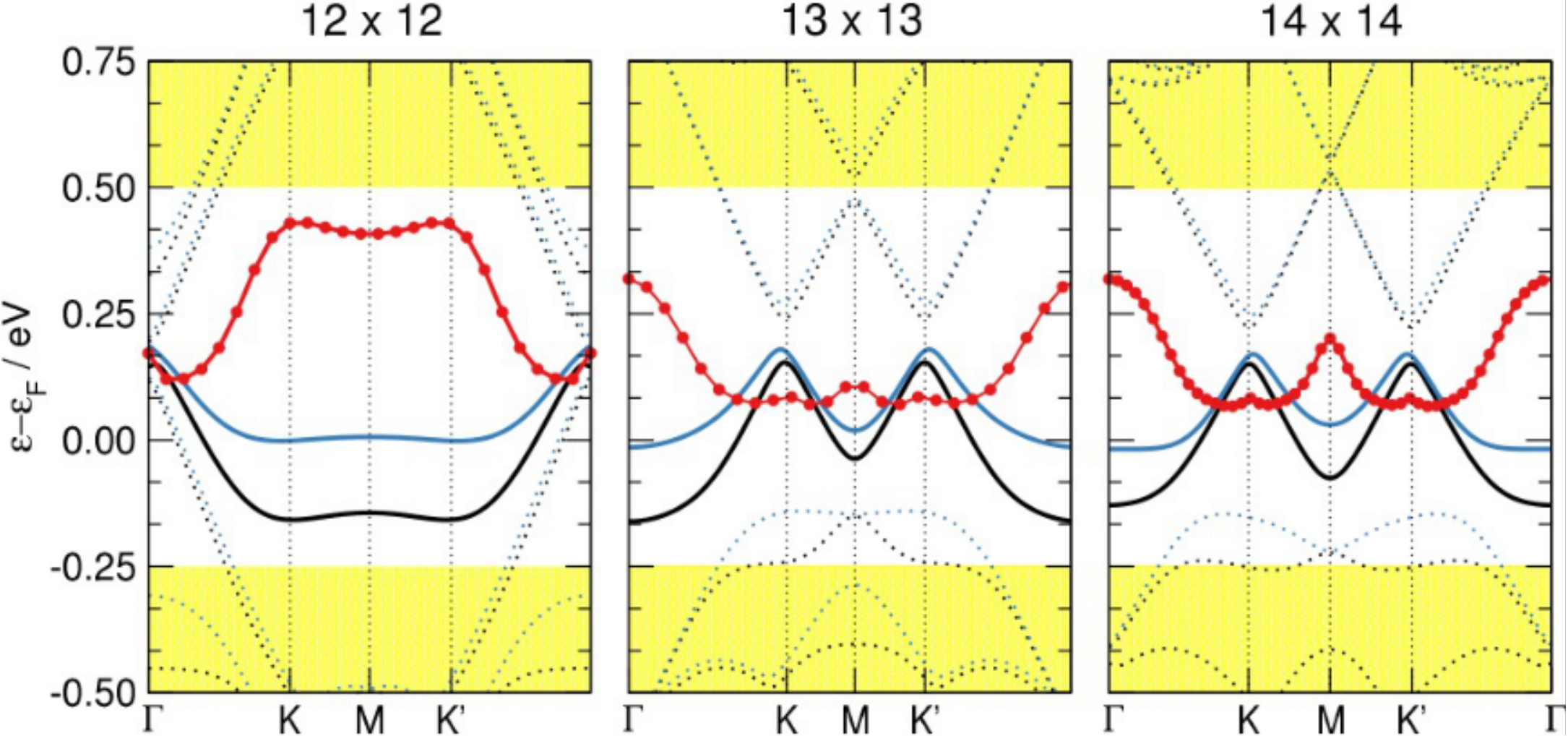}
\par\end{centering}
\caption{\label{fig:DFT low-energy bands}Spin-resolved low energy band structure
of the monovacancy in a 12x12, 13x13 and 14x14 supercell (as indicated).
The dispersive midgap band(s) are given as black and blue full lines
for the majority and the minority spin components, respectively. Also
shown as red lines the inverse participation ratios for the majority-spin
midgap state (magnified by 50 times).}
\end{figure}
\begin{figure}
\begin{centering}
\includegraphics[width=0.6\paperwidth]{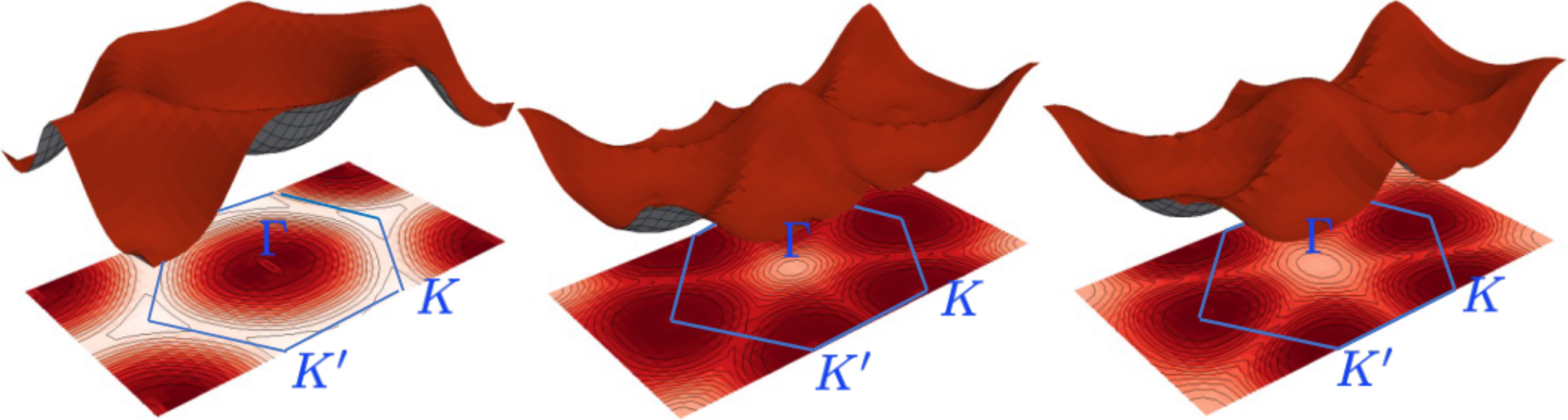}
\par\end{centering}
\caption{\label{fig:IPR maps}From left to right: behavior of the inverse participation
ratio of the majority-spin midgap state in the SBZ of the 12x12, 13x13
and 14x14 supercells with a C atom vacancy. The SBZ with its high-symmetry
points is given by the blue hexagons.}
\end{figure}
\begin{figure}
\begin{centering}
\includegraphics[width=0.4\columnwidth]{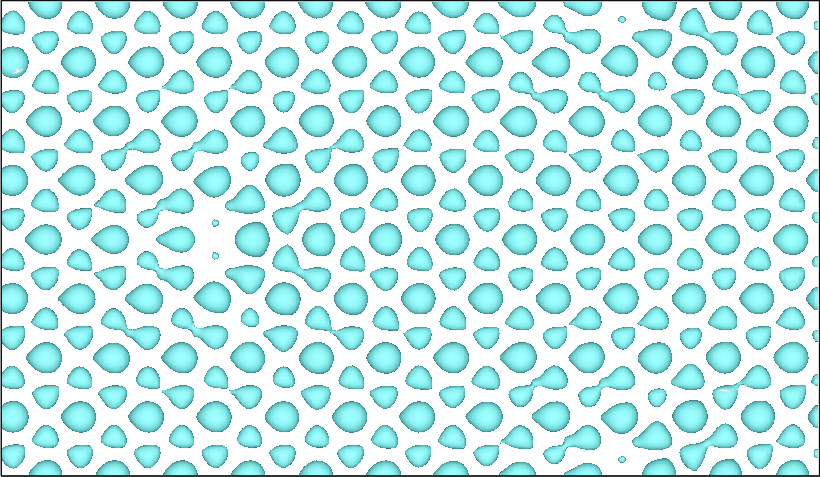}~~~~\includegraphics[width=0.4\columnwidth]{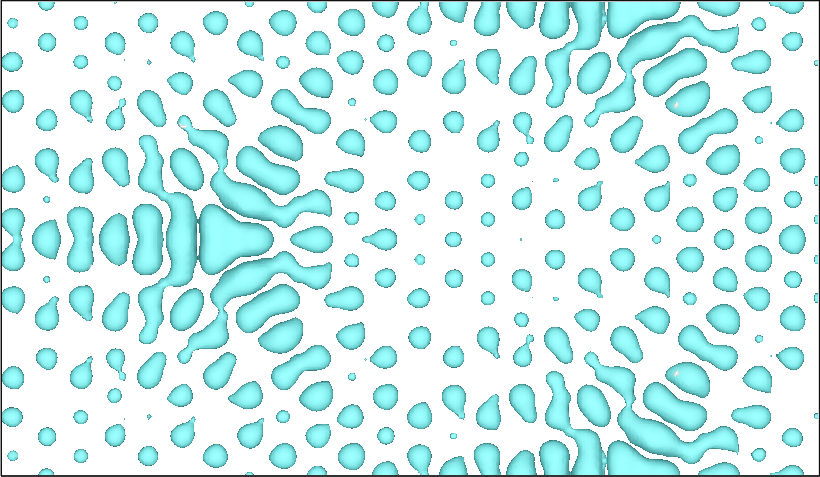}
\par\end{centering}
\caption{\label{fig:Density maps}Isocontours of the density $|\psi_{\mathbf{k}}(\mathbf{r})|^{2}$
for the $\Gamma$- and the $K$-point midgap-state wavefunctions (left
and right panel, respectively) in the 12x12 superlattice with a C
atom vacancy.}
\end{figure}
Fig. \ref{fig:DFT low-energy bands} reports the low-energy band-structure
of some $n\times n$ superlattices with large supercells ($n=12,13$
and $14$), featuring two dispersive midgap $\pi$ bands, one for
the majority spin species (black) and the other for the minority ones
(blue). Their width is large enough to compete with the Coulomb splitting
of the band, thereby determining a partial double occupation of the
band (hence, a partial quenching of the magnetization) in a way that
sensitively depends on the adopted supercell. To show that this is
indeed related to the anomalous delocalization discussed above, we
computed the IPR$_{q}$ (for $q=2$) of the \emph{first-principles}
$k$-wavefunctions describing the midgap state, namely the integrals
\begin{equation}
\text{IPR}_{q}(\mathbf{k})=\int_{U}|\psi_{\mathbf{k}}(\mathbf{r})|^{2q}d^{3}\mathbf{r}\label{eq:IPR (continuos)}
\end{equation}
where $U$ is the supelattice unit cell and $\psi_{\mathbf{k}}$ is
the (majority-spin) midgap state for $k-$vector $\mathbf{k}$. The
resulting IPR$_{2}$ is reported as red lines in Fig. \ref{fig:DFT low-energy bands}
and unambiguously shows the presence of delocalized states (small
values of the IPR) at the high-symmetry points of the SBZ where the
$K,K'$ states of pristine graphene fold to, i.e., the $\Gamma$ point
for $n=12$ and the $K,K'$ points otherwise. The behavior of the
IPR$_{2}$ over the SBZ is further shown in Fig. \ref{fig:IPR maps}
where it becomes clear that the anomalously delocalized states occupy
finite-sized regions of the SBZ (darker area in the color maps), which
are much larger than expected on the basis of the band plots alone.
Fig. \ref{fig:Density maps}, on the other hand, provides a real-space
picture of the wavefunction delocalization. Specifically, in that
figure we compare the anomalously delocalized density at the $\Gamma$
point of the 12x12 SBZ (left panel) with the ``normal'' (i.e., $1/r^{2}$-decaying)
counterpart at the $K$ point of the same SBZ (right panel). 

\begin{figure}
\begin{centering}
\includegraphics[width=0.6\paperwidth]{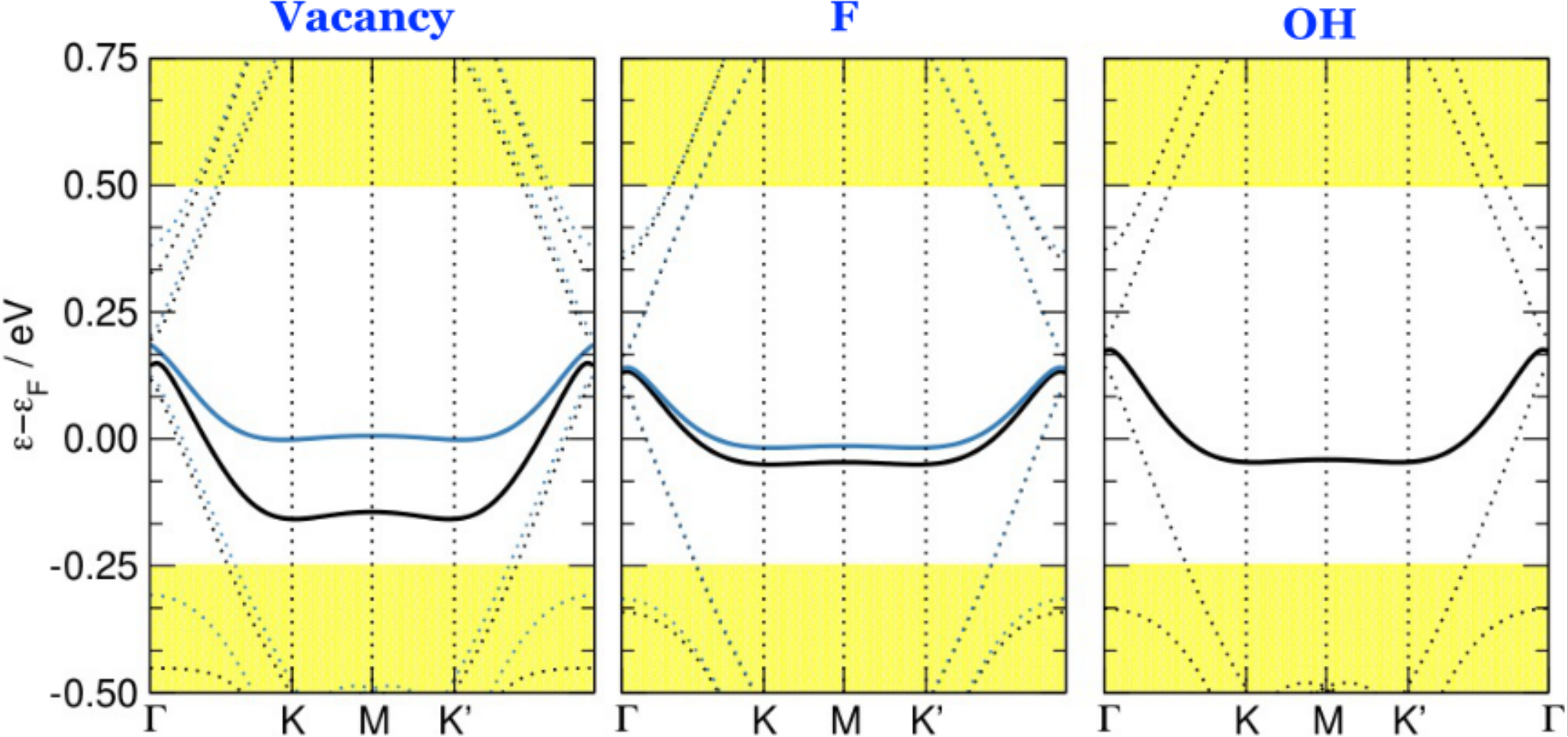}
\par\end{centering}
\caption{\label{fig:other defects}Spin-resolved low energy band structure
of the monovacancy and of the F and the OH adsorbates (as indicated),
as obtained in a 12x12 supercell. The dispersive midgap band(s) are
given as black and blue full lines for the majority and the minority
spin components, respectively. }
\end{figure}

Finally, we notice that the anomalous delocalization and the unphysical
dispersion is not limited to the C-atom vacancy, rather occurs for
any $p_{Z}$ defect, although its effects on the band filling and
the magnetization depends on the details of the defect (electronegativity).
This is shown in Fig. \ref{fig:other defects} for different $p_{Z}$
defects in the same 12x12 supercell, where the spin-splitting and
the band dispersion are seen to depend on the nature of the defect.
Hydrogen adatoms (not shown) are rather unique in this respect as
they are almost immune to unphysical dispersion. By this we do \emph{not}
mean that dispersion of the $\pi$ midgap band is absent for H adatoms,
rather that it is smaller than the spin-splitting of the band, hence
irrelevant for its occupation. See, for instance, Ref. \cite{Gonzalez-Herrero2016}
(Fig. S17) for the band structure of one H adatom in a large (and
skewed) supercell, which features a spin-splitting that is small on
an absolute scale but large enough to prevail over band dispersion.

\section{Discussion}

The above results unambiguosly show that modeling of a C-atom vacancy
in graphene in a periodic setting introduces anomalies in its electronic
structure that hamper the correct occupation of the bands and determine
a partial quenching of the magnetization. Although the difficulties
in assessing the magnetic moment with periodic \emph{first principles}
calculations have been acknowledged by several authors, they have
been attributed to the self-interaction error that plagues local and
semi-local functionals and that are known to seriously affect the
degree of electron localization / delocalization. In support to this
idea some authors have recently employed \emph{hybrid} functionals
in periodic calculations to show that they give a magnetic moment
that smoothly and rapidly converges to the expected value of $2$
$\mu_{B}$ when increasing the supercell size \cite{Valencia2017a,Ronchi2017}.
The use of these hybrid functionals that mix some fraction of exact
(Hartree-Fock) exchange to the semi-local one -- thereby partially
correcting the SIE --- has led to the immediate conclusion that the
SIE is crucial for the correct description of the vacancy and apparently
put an end to the long-debated issue of the magnetic properties of
an isolated C monovacancy in graphene. The conclusion however is \emph{incorrect},
since it is not physically sound and is inconsistent with experimental
findings. 

As mentioned in the Introduction, while the SIE argument might be
reasonable for a localized resonance characterized by a large Coulomb
``on-site'' energy, it is untenable for a semilocalized (not normalizable)
midgap state whose Coulomb energy hardly exceeds some tens of meV.
Indeed, carefully conducted experiments performed on defective graphene
samples have demostrated the presence of spin split peaks in the local
density of states in the neighborhoods of mono-vacancies, with energy
separations in the range 20-60 meV \cite{Zhang2016b}. These values
agree with the peak separation found by the authors of Ref. \cite{Gonzalez-Herrero2016}
in hydrogenated samples ($\sim$ 20 meV) for which (as seen above)
the involved resonant state is, for any practical aim, precisely the
same as that introduced by a C-atom vacancy. On the other other, the
spin-splitting produced by hybrid functionals is one order of magnitude
larger --- several hundreds of meV, depending on the supercell size
--- hence unphysical, except maybe in the true diluted limit. It
is thanks to the presence of this large splitting that one achieves,
with hybrid functionals, the correct occupation of the $\pi$ midgap
band and the smooth convergence of the magnetic moment. The electronic
structure, though, remains ``distorted'' and features yet an unphysically
dispersive $\pi$ midgap band, which is more than 0.5 eV wide for
the largest supercell considered so far (16 x 16, see Fig. 3 in Ref.
\cite{Ronchi2017}). This width is by no means comparable with the
``intrinsic'' width of the mono-vacancy which is experimentally
found of the order $\lesssim$ 0.05 eV \cite{Zhang2016b}. Hence,
while fixing the problem of filling the midgap band, hybrid functionals
do \emph{not} solve the core problem and, by providing a distorted
picture of the electronic structure, should be considered with caution
when investigating resonance states induced by $p_{Z}$-vacancies
in graphene. To be clear, local and semi-local functionals share similar
difficulties (after all, as shown above, these originate from the
one-electron part of the problem), but at least they describe the
spin splitting reasonably well (Fig. \ref{fig:DFT low-energy bands}).

\section{Summary and conclusions}

We have shown that $p_{Z}$ defects in graphene, when periodically
arranged, generate a $\pi$ ``midgap'' band whose real-space localization
properties change across the supercell Brillouin zone, from $1/r$
decaying expected for an isolated vacancy to fully delocalized. The
latter states are robust features inhereted from the pristine system
that spoil the correct description of an isolated vacancy. We emphasize
once again how they hamper the calculation of the magnitude of the
local magnetic moment of the monovacancy in graphene: in the presence
of inhomogeneities the (anomalous) delocalization broadens the defect-induced
resonace to such an extent that it spoils its filling. That is, the
actual width ($\Delta$) of the resonance is much larger than its
intrinsic one ($\Delta_{0}$) \footnote{According to Eq. \ref{eq:T-matrix}, this is proportional to $W^{2}\rho^{0}(\epsilon)$
at the resonant energy $\epsilon=\epsilon_{\text{res}}\approx0$. } and becomes comparable to (if not larger than) the spin- splitting
$U$ determined by the Coloumb repulsion in the resonant state ($\sim$
tens of meV). For the diluted limit of a single vacancy in the lattice
we should have $\Delta_{0}\ll U$, hence a net $\pi$ magnetic moment
of $1$ $\mu_{\text{B}}$ which adds to the one provided by the $\sigma$
state (1 $\mu_{\text{B}}$). However, in the large unit cell limit
of a periodic simulation one typically obtains $\Delta_{0}\ll\Delta\sim U$,
and the magnetization disappears (irrespective of the precise position
of the resonance relative to the Fermi level) unless $U$ is artificially
increased. The problem is not fixed by improving the theory level
--- e.g. by adding a fraction of exact exchange to alleviate the
self-interaction error that plauges popular density functionals ---
since it is of monoelectronic origin, as our TB results clearly show.
Its solution would require special sampling techniques in the slab-supercell
approach or, alternatively, the use of embedding techniques. 

\pagebreak{}


\providecommand{\newblock}{}

\end{document}